\documentclass[sigconf]{acmart}

\usepackage{booktabs} 

\usepackage{amssymb}
\usepackage{amsmath}
\usepackage{amsfonts}
\usepackage{graphicx}
\usepackage{url}
\usepackage{subfigure}
\usepackage{graphics}
\usepackage{color}
\usepackage{multirow}

\DeclareMathOperator*{\argmax}{arg\,max}
\DeclareMathOperator*{\argmin}{arg\,min}
\newcommand{\photo}{p}
\newcommand{\Photo}{\mathit{P}}
\newcommand{\hashtag}{h}
\newcommand{\Hashtag}{\mathit{H}}
\newcommand{\usr}{u}
\newcommand{\Usr}{\mathit{U}}
\newcommand{\loc}{\ell}
\newcommand{\Loc}{\mathit{L}}
\newcommand{\LocVar}{\mathcal{L}}
\newcommand{\timepoint}{t}
\newcommand{\Knowledge}{\mathit{K}}
\newcommand{\Adversary}{A}
\newcommand{\sme}[1]{\vec{m}_{#1}}
\newcommand{\utilityloss}[1]{\Delta_{#1}}

\copyrightyear{2018}
\acmYear{2018}
\setcopyright{iw3c2w3}
\acmConference[WWW 2018]{The 2018 Web Conference}{April 23--27, 2018}{Lyon,
France}
\acmBooktitle{WWW 2018: The 2018 Web Conference, April 23--27, 2018, Lyon, France}
\acmPrice{}

\acmDOI{10.475/123_4}
\acmISBN{123-4567-24-567/08/06}
\begin{document}
\title{Tagvisor: A Privacy Advisor for Sharing Hashtags}

\author{Yang Zhang}
\affiliation{
\institution{CISPA\\
Saarland Informatics Campus}
}
\author{Mathias Humbert}
\affiliation{
\institution{Swiss Data Science Center\\
ETH Zurich and EPFL}
}
\author{Tahleen Rahman}
\affiliation{
\institution{CISPA\\
Saarland Informatics Campus}
}
\author{Cheng-Te Li}
\affiliation{
\institution{Department of Statistics\\
National Cheng Kung University}
}
\author{Jun Pang}
\affiliation{
\institution{FSTC and SnT\\
University of Luxembourg}
}
\author{Michael Backes}
\affiliation{
\institution{CISPA Helmholtz Center i.G. \\
Saarland Informatics Campus}
}
\renewcommand{\shortauthors}{Y. Zhang et al.}

\begin{abstract}
Hashtag has emerged as a widely used concept of popular culture and campaigns, 
but its implications on people's privacy have not been investigated so far. 
In this paper, we present the first systematic analysis of privacy issues induced by hashtags. 
We concentrate in particular on location, which is recognized as one of the key privacy concerns in the Internet era. 
By relying on a random forest model, 
we show that we can infer a user's precise location from hashtags with accuracy of 70\% to 76\%, depending on the city.
To remedy this situation, we introduce a system called Tagvisor that systematically suggests alternative hashtags 
if the user-selected ones constitute a threat to location privacy. 
Tagvisor realizes this by means of three conceptually different obfuscation techniques 
and a semantics-based metric for measuring the consequent utility loss. 
Our findings show that obfuscating as little as two hashtags 
already provides a near-optimal trade-off between privacy and utility in our dataset. 
This in particular renders Tagvisor highly time-efficient, and thus, practical in real-world settings.
\end{abstract}


\maketitle

\section{Introduction}
\label{sec:introduction}

The development of information and communication technologies 
has progressively changed people's life style over the last decade.
Nowadays, people widely use online social networks (OSNs) 
and instant messaging to communicate with friends and maintain social relationships.
Besides, these new technologies have paved the way for concepts such as ``like'', ``follow'' and ``share'', 
which have been assimilated by OSN users within no time. 
One popular concept among these is the \emph{hashtag}. 
Defined as a word or non-spaced phrase preceded by the hash character \#, 
hashtag was created to serve as a metadata tag for people to efficiently search for information. 
Originally created on Twitter, hashtags have been widely adopted in many areas. 
Media campaigns are increasingly using hashtags to attract and engage customers (e.g., \#ShareACoke). 
Hashtags also enable people to stay updated on trending news stories. 
For instance, \#imwithher has been extensively used during the US 2016 election.

Hashtags have seen significant usage in online communication recently. 
Users have created many hashtags to convey meanings which previously did not exist in the English vocabulary. 
For instance, \#like4like indicates a promise of the users to like back the posts of those 
who like their post in OSNs~\cite{ZNHP17};
\#nofilter means the associated photo has been posted directly as taken, without being edited.
The large amount of hashtags provide us 
with an unprecedented chance to understand the modern society, 
and researchers are increasingly leveraging hashtags to conduct their study~\cite{SCFYCQA15,MAH16}. 

While bringing significant utility benefits, 
hashtags may carry sensitive private information about people using them. 
However, privacy threats arising out of hashtags have been largely overlooked. 
In this paper, we conduct the first study on addressing privacy raised by hashtags. 
Among all the private information, we concentrate in particular on mobility information, 
which is recognized as one of the key privacy concerns in the modern society~\cite{HSGH13,PZ15,BHMSH15,OHSHH17}. 
In fact, location arguably even constitutes the most sensitive information being collected by service providers~\cite{MHVB13} 
(e.g., being able to infer that a user is staying at a hospital can severely threaten his privacy), 
and it can be used to infer additional sensitive information such as friendship~\cite{SNM11,BHPZ17} 
and demographics~\cite{ZYZZX15,PZ17}.
A user can disclose his location information through many different ways such as sharing location in OSNs 
(often referred to as \emph{check-in})
and allowing mobile applications to collect their geo-coordinates through GPS sensors in the smartphone. 
Many users have recognized the inherent risks towards their location privacy and abstain from explicitly sharing their locations.
However, hashtags may also leak the exact location of a user, 
which is the primary focus of this paper. 

\smallskip
\noindent\textbf{Contributions.} 
Our contributions are two-fold: (i) a novel inference
attack for uncovering a user's fine-grained location based on posted hashtags, 
along with a comprehensive evaluation on a real-life OSN dataset, 
and (ii) a privacy-enhancing system, namely Tagvisor, 
that uses various obfuscation techniques for thwarting the aforementioned attack.

\smallskip
\noindent\emph{Attack.} 
Our inference attack aims at predicting the fine-grained location of 
a user's post given its hashtags 
by learning the associations between hashtags and locations.
The attack relies on a random forest classifier
and achieves remarkable accuracy.

We empirically evaluate our attack based on a comprehensive dataset of 239,000 Instagram posts 
with corresponding locations and hashtags 
from three of the largest English-speaking cities (New York, Los Angeles and London).
The locations we predict are the exact points of interest (POI) within each of the three cities, 
such as \emph{Land of Plenty}, a Chinese restaurant in New York.
The experiments show that our hashtag-based location-inference attack 
achieves an accuracy of 70\% in New York (for a number of around 500 considered locations) 
and 76\% in Los Angeles and London (for around 270 and 140 locations, respectively).
We also compare our classification method with other popular algorithms 
such as support vector machine and gradient boosting machine,
and show that our classifier outperforms them by at least 7\%. 
Furthermore, we empirically identify the number of hashtags maximizing the attack success to be seven.
Additionally, we evaluate our attack on a global level without prior knowledge of the city, 
and demonstrate that it can still reach an accuracy of more than 70\%. 
Finally, by considering two types of attackers, 
one with prior knowledge on the targeted users' hashtags and locations, 
and the other without, we show that the accuracy drops by around 20\% for the latter attacker.
This demonstrates that the adversary 
can enhance his model by learning per-user associations between hashtags and locations.

We stress that our work is not intended for users who intentionally use hashtags to reveal their locations, 
for example, when a user publishes a post with \#statueofliberty as one of the hashtags.
Instead, our work aims at helping users that might \emph{unintentionally} 
disclose where they are. 
The examples below, taken from the evaluation of the inference attack, illustrate this point further. 
A user's post shared with hashtags
\#music, \#musicphotography, \#soul, \#fujifilm, \#thephotoladies, and \#pancakesandwhiskey 
may not reveal anything significant about the location to most individuals. 
Our attack however correctly predicts that the user is at the \emph{Highline Ballroom}, 
a music venue in New York.
In another example from Los Angeles, a user sharing \#fighton \#trojans \#band \#ftfo
is correctly located by our attack at \emph{University of Southern California}. 

\smallskip
\noindent\emph{Defense.} 
To counter the aforementioned privacy violations, 
we develop Tagvisor: a system that provides recommendations, 
e.g., to an OSN user who wants to share hashtags without disclosing his location information. 
In particular, Tagvisor implements three different obfuscation-based mechanisms: 
hiding (a subset of) hashtags, replacing hashtags by semantically similar hashtags, 
and generalizing hashtags with higher-level semantic categories  (e.g., Starbucks into coffee shop). 
Tagvisor returns an optimal subset of obfuscated hashtags that, 
at the same time, guarantees some predefined level of location privacy and retains as much utility as possible.
To accurately quantify utility, we rely on an advanced natural language processing model, 
namely word2vec~\cite{MCCD13,MSCCD13}.

The empirical evaluation of our privacy-enhancing system shows that: 
(i) the replacement mechanism outperforms both hiding and generalization methods, 
(ii) the higher number of original hashtags, the better the utility for similar levels of privacy, 
and (iii) regardless of the original number of hashtags, 
obfuscating two hashtags provides the best trade-off between utility, privacy, and time efficiency. 
The latter finding notably demonstrates the practical feasibility of our privacy-preserving system 
given the computational capabilities of current mobile devices.

\smallskip
\noindent\textbf{Organization.} 
In Section~\ref{sec:model}, 
we present the  user and adversarial models we consider in this paper
as well as the privacy metrics for attack evaluation.
Our location inference attack 
is described in Section~\ref{sec:approach}.
We introduce our dataset in Section~\ref{sec:dataset} 
followed by the experimental evaluation in Section~\ref{sec:eval}.
We describe Tagvisor and the three obfuscation mechanisms 
for protecting users' location privacy in Section~\ref{sec:defense}. 
Section~\ref{sec:defeval} presents the performance of Tagvisor.
We provide a discussion in Section~\ref{sec:discussion}, 
address the related work in Section~\ref{sec:relwork}, 
and conclude the paper in Section~\ref{sec:conclu}.

\section{System Model}\label{sec:model}

In this section, we describe the user and adversarial models,
as well as the privacy metrics used throughout the paper. 

\subsection{User Model}

We use two sets $\Usr$ and $\Loc$ to represent users and locations, respectively.
A single user is denoted by $\usr$ and a location (POI) is denoted by $\loc$.
All users' posts are in set $\Photo$,
and a single post $\photo \in \Photo$ is defined as 
a quadruplet $\langle \usr$, $\loc$, $\timepoint$, $\Hashtag_\photo \rangle$:
semantically, it means $\usr$ shares the post $\photo$ at $\loc$ (checks in at $\loc$) at time $t$,
and $\Hashtag_\photo = \{\hashtag^1_\photo, \hashtag^2_\photo, \ldots, \hashtag^m_\photo\}$ 
contains all hashtags associated with the post.
In addition, $\Hashtag_\photo \subseteq \Hashtag_\loc \subseteq \Hashtag$ 
where $\Hashtag_\loc$ represents the entire set of hashtags shared (in posts) at location $\loc$ 
and $\Hashtag$ is the set of all hashtags in our dataset.

\subsection{Adversarial Model}\label{adv}

In this work, we consider an adversary who 
intends to infer the hidden location $\loc$ of a post $\photo$
by observing the set of hashtags $\Hashtag_\photo$ published with the post.
The adversary can be anyone who is able to crawl the data from OSNs 
such as Facebook, Twitter and Instagram.
Moreover, it could also model a service provider 
which does not have access to all its users' real locations.

In order to accurately infer hidden locations based on hashtags, 
the adversary needs to build an accurate knowledge $\Knowledge$ 
of associations between hashtags and locations. 
In this work, we assume he learns these associations 
based on previously disclosed posts that included both hashtags and location check-ins. 
This learning phase is explained in Section~\ref{sec:approach}. 
The associations can be learned in different ways. 
We consider here two different adversarial models. 
The first adversary, namely $\Adversary_1$, 
uses all the publicly shared posts he is able to collect, 
including those shared by the targeted users 
(users whose posts' locations the adversary intends to infer). 
In other words, in this model, 
we assume the targeted users have already shared some of their posts with both hashtags and check-ins. 
In the second model, the adversary ($\Adversary_2$) does not have access to any previous location-hashtag data shared by the targeted users. 
This models the case when the targeted users are particularly privacy-cautious~\cite{NZHP16}
and do not share any location with their posts. 
In this case, the adversary has to rely only on the location 
and hashtag data shared by other users to build his knowledge $\Knowledge$.

The output of our hashtag-based location inference attack on a post $\photo$ 
is the posterior probability distribution over all locations $\loc \in \Loc$ 
given the set of observed hashtags $\Hashtag_\photo$:
\begin{equation}
\Pr(\LocVar_\photo = \loc |  \Hashtag_\photo, \Knowledge),
\end{equation}
where $\LocVar_\photo$ is the random variable representing the location of post $\photo$.  
Based on this posterior distribution, the most likely location is:
\begin{equation}
\label{equ:mathias}
\loc_\photo^* = \argmax_{\loc \in \Loc} \Pr(\LocVar_\photo = \loc |  \Hashtag_\photo, \Knowledge).
\end{equation}
Equation~\ref{equ:mathias} is the core concept behind our inference attack, 
and we will describe its instantiation in detail in Section~\ref{sec:approach}.

\subsection{Privacy Metrics}\label{metrics}

We measure the location privacy of a user 
who shares a post $\photo$ and its corresponding hashtags $\Hashtag_\photo$ 
by using the expected estimation error, as proposed in~\cite{STBH11}. 
Formally, given the posterior probability distribution $\Pr(\LocVar_\photo = \loc |  \Hashtag_\photo, \Knowledge)$, 
the location privacy is defined as:
\begin{equation}
\sum_{\loc \in \Loc}{\Pr(\LocVar_\photo = \loc |  \Hashtag_\photo, \Knowledge) \cdot d(\loc, \loc^r_\photo)},
\label{equation:privacy}
\end{equation}
where $\loc^r_\photo$ is the real location of the user sharing the post $\photo$, 
and $d(\cdot, \cdot)$ denotes a distance function. 
If the geographical distance $d_g(\loc, \loc^r_\photo)$ is used, 
we will generally refer to this metric as the \emph{expected distance}.
In this paper, we rely on the haversine distance to measure the geographical distance.
If the binary distance, defined as
\begin{equation}
d_b(\loc, \loc^r_\photo)=
  \begin{cases}
  0 & \text{if } \loc = \loc^r_\photo\\
  1 & \text{otherwise}
  \end{cases}
\end{equation}
is used, 
Equation \ref{equation:privacy} is then equal to $1-\Pr(\LocVar_\photo = \loc^r_\photo |  \Hashtag_\photo, \Knowledge)$ 
and is referred to as the incorrectness~\cite{STBH11}. 
We also use the \emph{correctness} (to quantify the performance of the inference attack), 
which is the opposite of the incorrectness, 
and is simply equal to $\Pr(\LocVar_\photo = \loc^r_\photo |  \Hashtag_\photo, \Knowledge)$.
We further use the \emph{accuracy}, defined in the machine learning context, as another privacy metric.
Formally, accuracy is defined as $1 - d_b(\loc_\photo^*, \loc^r_\photo)$, 
where $d_b(\cdot, \cdot)$ is the binary distance defined above. 
We can alternatively use the inaccuracy, 
i.e., $d_b(\loc_\photo^*, \loc^r_\photo)$. 
Note that, depending on the context, the accuracy can refer to one sample or multiple. 
In the latter case, the accuracy represents in fact the \emph{average} accuracy over all samples (e.g., from a testing set).

\section{Inference Attack}
\label{sec:approach}
Our location inference attack relies on a random forest classifier. 
We encode the presence of a hashtag $h_i$ in a post $\photo$ as a binary value $x_\photo^i$, 
being equal to 1 if it is published with the post, and 0 otherwise.
Each post $\photo$ with its set of hashtags $\Hashtag_\photo$ at location $\loc$ 
can be represented by a feature vector $\vec{x}_\photo = (x_\photo^1, \ldots , x_\photo^n)$, where $n = |H|$, 
and by the label or class $y_\photo$, where $y_\photo = \loc_\photo^r$. 
The length of the feature vector is thus equal to the total number of unique hashtags in the entire dataset.

For training the random forest, 
the adversary uses as input the samples $ (\vec{x}_\photo ,y_\photo) \   \forall \photo \in \Photo_\textbf{train} $ 
where $\vec{x}_\photo = (x_\photo^1, \ldots , x_\photo^n)$
and $\Photo_\textbf{train} $ refers to posts in the training set 
that contains the auxiliary knowledge of the adversary. 
The adversary carries out his attack on the set $\Photo_\textbf{tar}$ 
by trying to classify each sample post $p \in \Photo_\textbf{tar}$ 
using the features $\vec{x}_\photo$ on the trained forest.
Among several ways of obtaining class probabilities 
in an ensemble of decision trees~\cite{B07}, 
we employ averaging of the votes of all the trees of the forest. 
The total number of trees that vote for each class is divided 
by the total number of trees 
to obtain the posterior probability distribution 
$\Pr(\LocVar_\photo = \loc |  \Hashtag_\photo, \Knowledge) \  \forall \loc \in \Loc$. 
This outcome is used to quantify privacy with the metrics defined in Section \ref{metrics}.
  
\section{Dataset}
\label{sec:dataset}

\begin{figure}[!t]
\centering
\includegraphics[width=0.5\columnwidth]{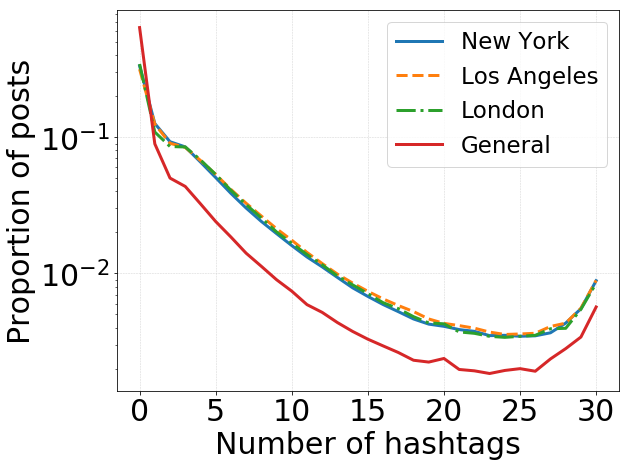}
\caption{Distribution of the proportion of posts with a certain number of hashtag(s), from 0 to 30 hashtags.}
\label{fig:hash_dist}
\end{figure}

\begin{table}
\centering
\caption{Pre-processed data statistics. \label{table:pro_dataset}}
\scalebox{0.9}{
\begin{tabular}{l|c|c|c}
\hline
& New York & Los Angeles & London\\
\hline
No. of posts & 144,263 & 61,767 & 34,018\\
No. of hashtags & 8,552 & 4,600 & 2,395\\
No. of users  & 3,911& 1,625 & 992\\
No. of locations  & 498 & 268 & 141\\
\hline
\end{tabular}
}
\end{table}

We collect our experimental data from Instagram,
one of the most popular OSNs and
the major platform for hashtag sharing.
Besides, Instagram users have been shown to share their locations 
an order of magnitude more often than in Twitter~\cite{MHK14},
and Instagram's localization function was linked with Foursquare,
which enabled us to enrich each location with name and category,
not just geograpical coordinates as in many existing datasets~\cite{CML11}.
The category information is in fact a building block 
for one of our defense mechanisms, namely the semantic generalization (Section~\ref{sec:defense}).

\begin{table*}[!t]
\centering
\caption{Performance of location inference across all the cities for different adversary models and baseline. \label{table:performance}}
\scalebox{0.9}{
\begin{tabular}{l|c|c|c| c|c|c |c|c|c |c|c|c}
\hline
&\multicolumn{3}{c|}{New York} & \multicolumn{3}{c|}{Los Angeles} & \multicolumn{3}{c|}{London} & \multicolumn{3}{c}{All cities}\\
\hline
& $\Adversary_1$ &$\Adversary_2$ & baseline & $\Adversary_1$  & $\Adversary_2$ & baseline & $\Adversary_1$  & $\Adversary_2$ & baseline & $\Adversary_1$  & $\Adversary_2$ & baseline\\
\hline
Correctness & 0.613 & 0.468 & 0.015 &  0.685 & 0.502 & 0.015 & 0.686 & 0.552 & 0.020 & 0.624  & 0.465  & 0.010 \\
Expected distance (km) & 0.917 & 1.272  & 4.198 & 1.870 & 3.046 & 11.275 & 0.857 & 1.575 & 4.518 & 211.471 & 345.980  & 3563.082\\
Accuracy & 0.697 & 0.556 & 0.053 & 0.758 & 0.597 & 0.048 & 0.761 & 0.617 & 0.051 & 0.712  & 0.560 & 0.045  \\
\hline
\end{tabular}
}
\end{table*}

Our data collection was conducted in January 2016. 
We concentrate on three of the largest cities in the English-speaking world, namely
New York, Los Angeles and London.
In the first step, we rely on the Foursquare API to crawl
all the location IDs in the three cities,
and extract these locations' names and categories.
Then, we use the function \texttt{/locations/search} provided by Instagram's public API 
to map the previously obtained Foursquare's location IDs
to their corresponding Instagram's location IDs.
Finally, for each Instagram location, we 
collect its publicly available posts' metadata (user ID, time and hashtags) 
in the second half of 2015 (2015.7.1-2015.12.31).

As in~\cite{CML11,SNM11,ZP15,PZ17b,BHPZ17},
we perform the following filtering operations on our collected data.

\begin{itemize}
\item We filter out user accounts whose number of followers 
are above the 90th percentile (celebrities) or under the 10th percentile (bots).
To address the data sparseness issue, 
we also filter out users with less than 20 check-ins in each city.
\item We filter out hashtags appearing less than 20 times in each city,
following a classical filtering strategy used 
by the natural language processing community.
\item For each city, we concentrate on locations with at least 50 check-ins. 
This helps us filter out the locations that are rarely visited by people 
while still ensuring that we are not only concentrating on the most famous places.
Knowing a user being at the considered locations still threatens his privacy to a large extent. 
Additionally, from a supervised learning point of view, 
we need sufficient data for each class (each location) to train a robust classifier, such that a meaningful attack can be conducted.
\end{itemize}

After the filtering, we obtain 144,263 posts in New York, 61,767 posts in Los Angeles 
and 34,018 posts in London (Table~\ref{table:pro_dataset}). 
Despite the filtering strategies, we are still left with a large dataset 
with substantially more users and fine-grained locations 
compared to previous works.
For instance, the dataset used in~\cite{STBH11} 
contains 20 users' mobility data in a $5 \times 8 $ grid over the city of San Francisco. 

Figure~\ref{fig:hash_dist} displays the distribution of the number of hashtags per post over our entire dataset. 
For all three cities, there are around 40\% of posts without any single hashtag, 
between 11 and 14\% with one hashtag, 8-9\% with two hashtags, and 8\% with three hashtags. 
The general curve (in red) represents the distribution of number of hashtags for posts without location check-in. 
The latter dataset is collected 
following the same methodology as in~\cite{SCFYCQA15}.
We randomly sample over 10 million user IDs on Instagram and collect all their published posts' hashtags.
In the general set of posts, there are 46\% of the posts that include at least one hashtag. 
This shows that, almost half of the posts could be targeted by our hashtag-based location inference attack. 
The similar trend between the general curve and those with location information shows 
that users who do not reveal their location may nevertheless disclose hashtags
that could then be used by an adversary to infer their locations. 
As expected, the distribution drops quite fast with an increasing number of hashtags, 
for both datasets with and without location information. 
The small increase close to 30 is due to Instagram's upper bound of 30 hashtags per post.

\section{Attack Evaluation}
\label{sec:eval}
In this section, we present the evaluation results of our location inference attack. 

\subsection{Experimental Setup}
As described in Section~\ref{sec:approach},
our attack relies on a random forest classifier, 
and we set the number of trees for the random forest to be 100 following common practice~\cite{OPB12}.

To comply with the different background knowledge of $\Adversary_1$ and $\Adversary_2$,
we split the dataset as the following.
For adversary $\Adversary_1$, 
we randomly assign 80\% of the posts for training and 20\% for testing. 
For adversary $\Adversary_2$, 
we split all users randomly: 
posts from 80\% of the users are put into the training set
and posts from the other 20\% are in the testing set.
The random split is repeated 10 times, and we report the average results.

\subsection{General Inference Results}

\noindent\textbf{Adversaries.}
We present the performance of our location inference attack in Table~\ref{table:performance}.
In each city, our attack achieves a strong prediction.
For instance, the accuracy is around 70\% for the New York dataset under adversary $\Adversary_1$, 
which indicates the attacker is able to predict the exact locations (out of 498)
of 20,196 posts (out of 28,852 testing posts).
The small expected distance in all cities further indicates that even when the prediction is wrong,
the attacker is still able to narrow down the target's location into a small area.
The performance of adversary $\Adversary_2$, however, drops in all the cities. 
This is because adversary $\Adversary_2$ has no prior knowledge 
of the hashtag-location association of the targeted users. 
Nevertheless, adversary $\Adversary_2$ still achieves 
a relatively high prediction success (e.g., the correctness is above 0.55 in London) 
showing that learning per-user associations between hashtags and locations is helpful but not absolutely needed.

\smallskip
\noindent\textbf{Cities.}
Both adversaries achieve the strongest prediction in London, 
followed by Los Angeles and New York. 
The reasons are manifold, ranging from cultural differences to hashtag usage. 
We notice that New York has the largest number of locations (498), 
i.e., the highest number of classes for the random forest model, 
thus making the classification the most difficult. 
Meanwhile, even though both adversaries in Los Angeles 
achieve the second highest accuracy and correctness, 
they perform the worst in the expected distance metric. 
This is due to the different densities of the three cities:
Los Angeles covers a larger area with places being more uniformly distributed in the geographical space
than New York and London.

\smallskip
\noindent\textbf{Baseline.}
To demonstrate that our high prediction performance 
is not due to the skewed distribution of the location check-ins, 
we further establish a baseline model that relies only on the locations' distribution (in the training set) 
to predict a targeted user's location.  
The adversary infers that the user is at the most likely location, 
i.e., the location with most check-ins in the training set, independently of the hashtags. 
As depicted in Table~\ref{table:performance}, 
both $\Adversary_1$ and $\Adversary_2$ achieve at least a 10-time higher accuracy than the baseline, 
and a 27-time higher correctness than the baseline. 
We observe similar results when the adversary targets all cities at the same time. 
As for the expected distance metric, the baseline is outperformed by 3 to 5 times 
depending on the considered adversary and city. 

\smallskip
\noindent\textbf{Other classifiers.}
Besides random forest,
we further experiment with two other classifiers:
gradient boosting machine (GBM)
and support vector machine (SVM).
For GBM, we observe that 100 trees provide the best accuracy,
and for SVM, we use the radial basis function kernel.
GBM's accuracy is around 7\% worse than random forest in all the cities, 
and SVM achieves the worst performance with accuracy in London of only 0.15.
This suggests that random forest is the most suitable algorithm for learning the association between hashtags and locations.

\subsection{Privacy vs. Number of Hashtags}\label{privvsnum}

Next, we study the relation between the number of shared hashtags and
the prediction performance, i.e., 
how many hashtags are necessary to determine where a post is shared. 
To this end, we group the posts in the testing set by the number of hashtags 
they contain and compute the correctness for each group.

\begin{figure}[t]
\centering
\subfigure[]{
\includegraphics[width=0.46\columnwidth]{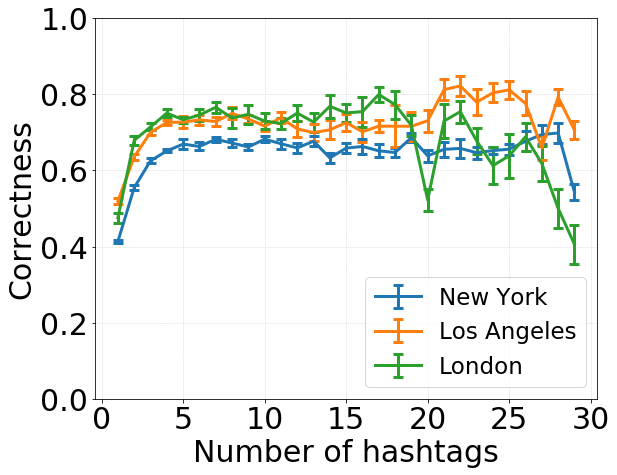}\label{fig:correctness_a1}}
\subfigure[]{
\includegraphics[width=0.46\columnwidth]{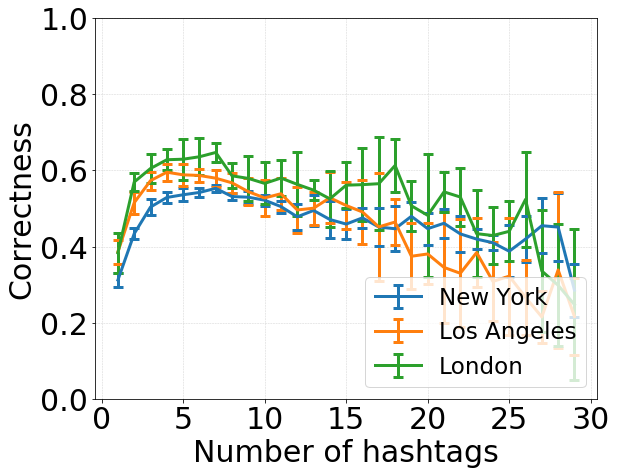}\label{fig:correctness_a2}}
\caption{Correctness of adversary \subref{fig:correctness_a1} $\Adversary_1$ and \subref{fig:correctness_a2} $\Adversary_2$
with respect to the number of hashtags shared in posts. \label{figure:cnt_correctness}}
\end{figure}

Figure~\ref{figure:cnt_correctness} depicts 
the average correctness for the two adversary models 
(results for accuracy follow a very similar trend). 
As we can see, for posts appended with only one hashtag, 
the prediction performances are relatively weak in all the cases 
since one hashtag does not provide enough information. 
However, when the number of hashtags increases to two, 
we observe significant increases under both adversary models. 
Especially for $\Adversary_1$, the correctness is increased 
by more than 40\% for all the cities. 
This shows that hashtags contain highly location sensitive information. 
The performance of both adversary models 
is strongest when the targeted post contain around 7 hashtags. 
Beyond this number, we observe an interesting difference. 
On one hand, the average correctness of adversary $\Adversary_1$ 
remains stable for posts with more hashtags.  
We can observe a small decrease only for London 
due the relatively smaller size of testing data which probably creates more noisy results. 
On the other hand, the performance of adversary $\Adversary_2$ decreases for all cities 
when it faces posts with increasing number of hashtags. 
This happens because, by including more hashtags, 
a user confuses the classifier which has no information about the past activity of the user, 
as $\Adversary_2$ has not built user-specific associations between hashtags and locations. 
Therefore, a user who has never shared any locations (the assumption of $\Adversary_2$) 
is less vulnerable to the attack. 
This, along with the conclusions drawn from Table~\ref{table:performance} 
that $\Adversary_2$ performs worse than $\Adversary_1$ in general, 
further suggests that a cautious user is always rewarded in matters of privacy.

\subsection{Privacy across All the Cities}
So far, our inference attack is conducted at the city level. 
However, it is also interesting to see 
whether our attack can be generalized to the global level:
Given a post with hashtags,
can we predict where it is published among all the locations in New York, Los Angeles and London.
To this end, we combine the datasets from the three cities together
and perform the aforementioned split with respect to the two adversary models 
for training and testing. 
As shown in Table~\ref{table:performance}, 
the performance of the global level attack drops compared to the city level attack, 
especially for London and Los Angeles. 
However, the attack is still rather effective,
e.g, adversary $\Adversary_1$ achieves 0.712 accuracy and 0.624 correctness. 
Meanwhile, the expected distance grows much larger, 
which is mainly due to the misclassification among different cities. 
However, considering the actual distances between the three cities, 
this expected distance is still very small,
and both $\Adversary_1$ and $\Adversary_2$ perform much better than the baseline model.

Through the above experiments, we have demonstrated the severe privacy threat carried by hashtags. 
In the next section, we propose a first of its kind privacy advisor which we coin Tagvisor, 
that uses three defensive mechanisms to provide users with advice when sharing hashtags.

\section{Countermeasures}
\label{sec:defense}
In this section, we present \emph{Tagvisor}, 
a privacy advisor that is based on different obfuscation mechanisms 
for enhancing the user's privacy and preserving as much semantic utility as possible. 
We first introduce the different components of Tagvisor.
Then, we describe our approach for measuring hashtags' utility. 
Finally, we present the three obfuscation mechanisms. 

\subsection{Tagvisor}\label{cons}
Tagvisor resides on the user's device and, 
whenever the user wishes to share a post with some hashtags without revealing his location, 
Tagvisor tests whether its classifier can correctly predict the post's location,
i.e., whether the user's location privacy is compromised. 
If so, Tagvisor suggests alternate sets of hashtags that 
result in an enhancement of privacy without significant utility loss. 
The system can also suggest the optimal set of hashtags 
that minimizes the utility loss while providing a certain level of location privacy.

Formally, the Tagvisor's defense mechanism can be interpreted as an optimization problem.
For each new post $p$ that a user wants to share with an original set of hashtags $\Hashtag_\photo$, 
Tagvisor suggests the optimal set of hashtags $\Hashtag_\photo^*$ as follows:
\begin{equation}\label{equa:utility}
\Hashtag_\photo^* = \argmin_{\Hashtag_\photo' \in \mathcal{H}_\photo}
 \utilityloss{\Hashtag_\photo, \Hashtag_\photo'}\ \ \ \text{ subject to }\ \ \ LP_{\Hashtag_\photo'} \geq \alpha.
\end{equation}
Here, $\utilityloss{\Hashtag_\photo, \Hashtag_\photo'}$ represents the utility loss, 
quantified by the semantic distance between $\Hashtag_\photo$ and $\Hashtag_\photo'$. 
$LP_{\Hashtag_\photo'}$ represents the privacy level for the set $\Hashtag_\photo'$, 
which can be quantified with any of the metrics defined in Section~\ref{sec:model}, 
and $\alpha$ is the minimal privacy level desired by the user. 
For example, a simple and intuitive profile would be to set $\alpha = 1$ 
with $LP$ measured by $d_b(\loc_\photo^*, \loc^r_\photo)$, the inaccuracy. 
This requirement ensures that the user only shares hashtags 
that do not enable the attacker to infer his exact location $\loc^r_\photo$.

$ \mathcal{H}_\photo$ can be as big as the power set of $H$. 
In order to keep complexity at a reasonable level, 
we can restrain it to a smaller subset depending 
on the different protection mechanisms 
and the original set of hashtags $\Hashtag_\photo$. 
The user can also set a maximum number of hashtags 
that the privacy mechanism can modify. 
We evaluate these more practical constraints in Section~\ref{sec:defeval}.

\begin{figure}[t]
\centering
\includegraphics[width=0.71\columnwidth]{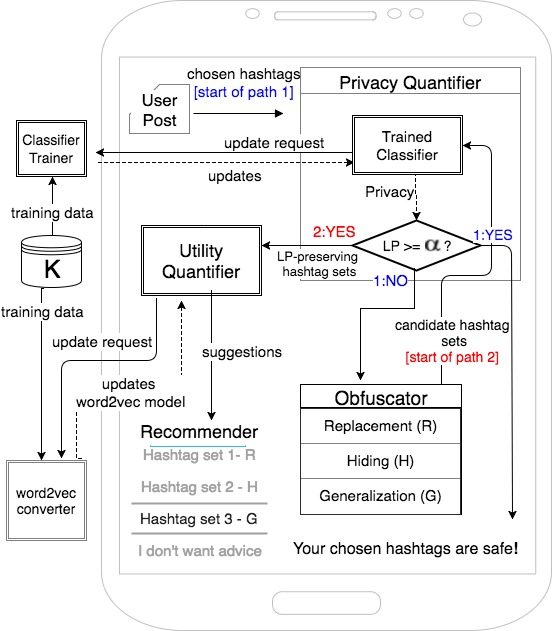}
\caption{Main building blocks of Tagvisor \label{fig:tagvisor}}
\end{figure}

Tagvisor consists of four main blocks.
Figure \ref{fig:tagvisor} shows the main components of the Tagvisor system. 
R, H and G in Recommender represents obfuscation by Replacement, Hiding and Generalization, respectively. 
If the privacy is preserved with the original set of hashtags chosen by the user, 
they are directly published (1:YES). 
Otherwise, they go to the obfuscator (1:NO) which then sends the modified sets to the privacy quantifier. 

\smallskip
\noindent\textbf{Privacy quantifier.}
The \emph{privacy quantifier} contains the \emph{trained classifier}, 
i.e., our random forest model, and a \emph{condition checker}. 
The \emph{trained classifier} is regularly updated with new knowledge from the external \emph{classifier trainer}. 
This classifier trainer uses the data $K$ to constantly improve its random forest model. 
It could be running on the user's personal computer, 
or deployed by a non-governmental organization such as the Electronic Frontier Foundation.
The \emph{condition checker} checks 
whether the level of location privacy of the user (using the classifier output) 
is above the threshold $\alpha$. 
If not, the user chosen hashtags are sent to the \emph{obfuscator} for determining alternative hashtag sets. 
Among these suggested sets of hashtags, 
those for which the privacy condition in the \emph{condition checker} 
is satisfied are then fed to the \emph{utility quantifier}.

\smallskip
\noindent\textbf{Obfuscator.}
The \emph{obfuscator} consists of three components 
that represent the different obfuscation mechanisms: \emph{hiding, replacement} and \emph{generalization}. 
Each component produces a set of hashtags 
by employing the corresponding defense mechanism. 
These hashtag sets are fed to the \emph{privacy quantifier} 
which filters out those that do not meet the required privacy threshold $\alpha$ (2:YES in Figure~\ref{fig:tagvisor}). 
 
\smallskip
\noindent\textbf{Utility quantifier.}
The \emph{utility quantifier} also resides on the user's device. 
It maintains records of the utility model and periodically updates itself 
by querying the external \emph{word2vec converter} 
(which continuously updates itself with any new data from the knowledge database $K$). 
As will be described in Subsection~\ref{subsec:util}, 
the \emph{utility quantifier} uses the semantic representation of each hashtag 
in the set of input hashtags that it receives from the \emph{privacy quantifier} 
to calculate their utility with respect to the original set of hashtags. 
Finally, it returns the hashtag set(s) providing minimal utility loss to the \emph{recommender}. 
Depending on the user's preferences, 
it can return one recommended hashtag set per obfuscation method 
(as shown in Figure~\ref{fig:tagvisor}), or only one optimal hashtag set for all methods. 
 
\smallskip
\noindent\textbf{Recommender.}
The \emph{recommender} suggests, 
for each obfuscation method, 
location-privacy-preserving hashtag sets 
having minimum semantic distance to the original hashtags chosen by the user. 
It can also suggest the optimal hashtag set among all methods, 
if the user does not want to make a decision by himself. 
It could also provide additional recommendations for a desired obfuscation method 
and a desired maximum number of hashtags to be obfuscated.

\subsection{Utility Metric}\label{subsec:util}

The defense mechanisms should yield hashtags
that retain the semantics of the original hashtags
to the largest possible extent,
while still defeating the attacker
by misleading the random forest classifier. 
Removal of all hashtags, for example, reduces the attacker to a random guesser. 
However, for the user, the consequence is a complete loss of utility. 

We rely on the semantic distance between the original set of hashtags 
and the sanitized set to quantify the corresponding utility loss.
We capture the semantic meaning of each hashtag, and set of hashtags, 
by using \emph{word2vec}, the state-of-the-art method for representing language semantics~\cite{MCCD13,MSCCD13}.
word2vec maps each hashtag
into a continuous vector space of $d$ dimension,\footnote{Following the original works~\cite{MCCD13,MSCCD13},
$d$ is set to 100 in our experiments.} 
i.e.,
\begin{equation}
{\it word2vec}: \Hashtag \rightarrow \mathbb{R}^d,
\end{equation}
using a (shallow) neural network with one hidden layer.
The objective function of word2vec is designed to preserve
each hashtag's context, i.e., neighboring hashtags.
Therefore, if two hashtags are often shared together in posts,
they will be close to each other in the learned vector space.
The concept that word2vec can represent a word's semantic meaning
originates from the distributional hypothesis in linguistics,
which states that words with similar contexts have similar meanings.
For example, in our dataset, \#travel and \#tourist are semantically close
under word2vec. We define the word2vec representation of a hashtag $h$ as $\vec{v}_\hashtag = word2vec(h)$.

We train our word2vec model on the whole set of posts available to obtain a semantic vector for each hashtag in $\Hashtag$.
Following previous works~\cite{YHBP14,KR15}, 
we express the semantic meaning of hashtags  $\Hashtag_\photo$ in a post $\photo$ as the average of their semantic vectors:
\begin{equation}
\sme{\Hashtag_\photo} = \frac{\sum_{\hashtag\in\Hashtag_\photo} \vec{v}_\hashtag}{\vert\Hashtag_\photo \vert}.
\end{equation}
Then, the utility loss of replacing the original set $\Hashtag_\photo$ by $\Hashtag'_\photo$
is measured as the Euclidean distance between $\sme{\Hashtag_\photo}$ and $\sme{\Hashtag'_\photo}$:
\begin{equation}
\utilityloss{\Hashtag_\photo, \Hashtag'_\photo} = \vert\vert \sme{\Hashtag_\photo} - \sme{\Hashtag'_\photo}\vert\vert.\label{eq:loss}
\end{equation}
We use Euclidean distance 
since it is the most common method to measure distances 
in the word2vec generated vector space~\cite{MCCD13,MSCCD13}. 

\begin{figure*}[h]
	\centering
	\subfigure[]{\includegraphics[width=0.57\columnwidth]{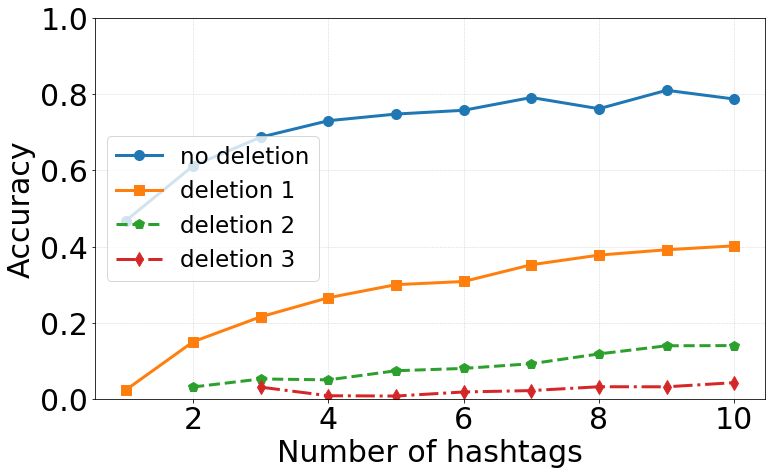}\label{fig:del_accuracy}}
	\hspace{5mm}
    \subfigure[]{\includegraphics[width=0.57\columnwidth]{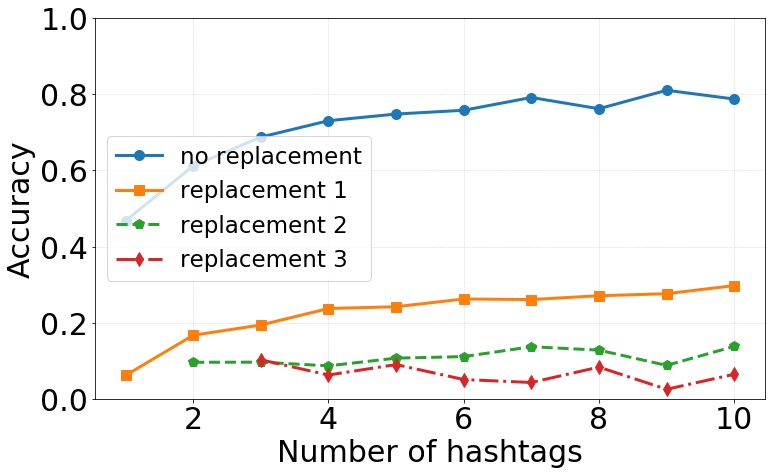}\label{fig:rep_accuracy}}
	\hspace{5mm}
    \subfigure[]{\includegraphics[width=0.57\columnwidth]{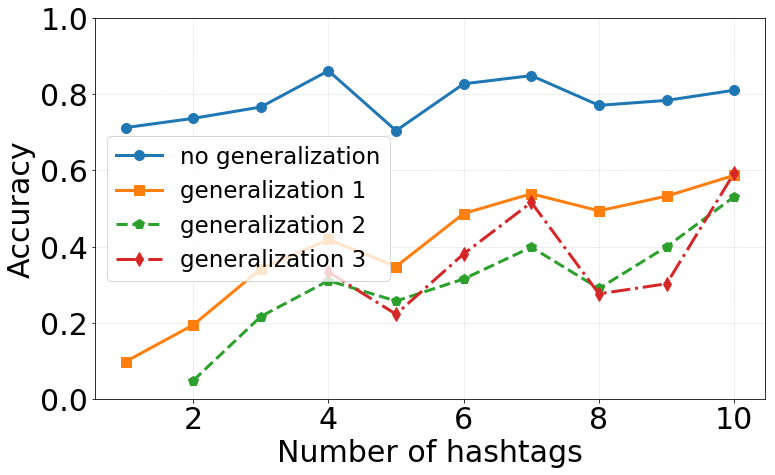}\label{fig:gen_accuracy}}
    \caption{Evolution of the accuracy ($\Adversary_1$) with respect to 
    different original numbers of hashtags to be shared (x-axis) 
    and numbers of hashtags to be obfuscated (from 0 to 3) 
    for (a) hiding, (b) replacement, and (c) generalization in New York.}
	\label{fig:accuracy_obfuscation}
\end{figure*}

\subsection{Obfuscation Mechanisms}

We now describe the three obfuscation mechanisms used by Tagvisor for protecting location privacy. 

\smallskip
\noindent\textbf{Hiding.}
Hiding (also referred to as deletion)
simply suggests a subset of the original hashtags $\Hashtag_\photo$ chosen by the user. 
There are in total $2^{|\Hashtag_\photo|}-1$ such subsets. 
One can limit the number of hashtags to be removed by the threshold $t_h$ to optimize the running time, or utility. 
All generated subsets of hashtags are sent to the condition checker to verify 
whether they satisfy the location privacy constraint. 
The subsets providing enough privacy are sent to the utility quantifier 
that picks the one that minimizes utility loss.

\smallskip
\noindent\textbf{Replacement.}
This mechanism replaces the original hashtags 
by others in the set of hashtags $\Hashtag$ to mislead the adversary. 
In order to keep the search complexity reasonable, 
we have to restrain the set of potential hashtags 
to replace each original hashtag with. 
We fix a threshold $t_s$ and focus on the $t_s$ hashtags 
that are semantically closest to the original hashtag, as defined in Equation \ref{eq:loss}. 
This ensures that the set of candidate hashtags minimizes utility loss. 
This bounds the search space from above by $(t_s+1)^{|\Hashtag_\photo|}-1$. 
As in the hiding mechanism, one can further reduce the time complexity 
by bounding the number of hashtags 
to be replaced by a threshold $t_r$ similar to $t_h$. 
The generated subsets are eventually sent to the privacy condition checker and the utility quantifier.

\smallskip
\noindent\textbf{Generalization.}
This mechanism generalizes each original hashtag 
by one representing a semantically broader category. 
Our evaluation only focuses on hashtags corresponding to locations in Foursquare. 
Every location identifier in Foursquare is mapped to a category identifier at two levels. 
For example, \emph{Harrods} has the category \emph{department store} at the second, lower, level 
and category \emph{shop} at the first, higher, level. 
Since not all hashtags are generalizable (e.g., \#love or \#instagood), 
we represent the subset of generalizable hashtags in a given post 
as $\Hashtag_g \subseteq \Hashtag_p$. 
The search space is then equal to $3^{\Hashtag_g}-1$. 
To reduce the time complexity, 
we can also fix a threshold $t_g$ of the maximum number of hashtags to be generalized.

\section{Defense Evaluation}\label{sec:defeval}

In this section, we present the evaluation results of the defense mechanisms 
presented in the previous section. 
We concentrate on the results for the strongest adversary, $\Adversary_1$, in New York 
since it includes the maximum number of users and locations.
The results for Los Angeles and London follow a similar trend, 
which we skip due to space constraints.
We consider $t_s = 2$ for the replacement mechanism in order to reduce the search space, 
and thus improve the time efficiency. 
The other parameters $t_h$, $t_r$ and $t_g$ are left free and their impact is evaluated in our experiments.

We begin with the impact of obfuscating hashtags on the attack accuracy (thus privacy)
with respect to each obfuscation mechanism. 
Then, we compare the utility (corresponding to the obtained privacy level) 
between the hashtags suggested by
the three obfuscation mechanisms separately and 
the optimal global solution. 
Finally, we evaluate the time efficiency of Tagvisor.

\subsection{Accuracy vs. Obfuscation Level}

Figure~\ref{fig:accuracy_obfuscation} shows the impact of 
(a) hiding, (b) replacement, and (c) generalization on the attack accuracy in New York.
Except for generalization, we notice that the larger the number of hashtags obfuscated, 
the lower the average accuracy (different curves), thus the higher the privacy, as expected. 
We additionally observe that, 
even with 10 original number of hashtags, 
the average accuracy with two obfuscated hashtags only is already very low (smaller than 0.2). 
This means that the attacker is able to correctly infer the location 
of only less than 20\% of the sample posts. 
These results demonstrate the effectiveness of the hiding and replacement mechanisms. 
However, generalization (Figure~\ref{fig:gen_accuracy}) 
cannot bound the accuracy of the attack 
when the number of hashtags to be shared increases.
Note that the curves are less smooth
in this latter case because we have more constraint on our testing set: 
our testing set is much smaller for this scenario since not all hashtags are generalizable.

\begin{figure*}[h]
\centering
\subfigure[]{\includegraphics[width=0.57\columnwidth]{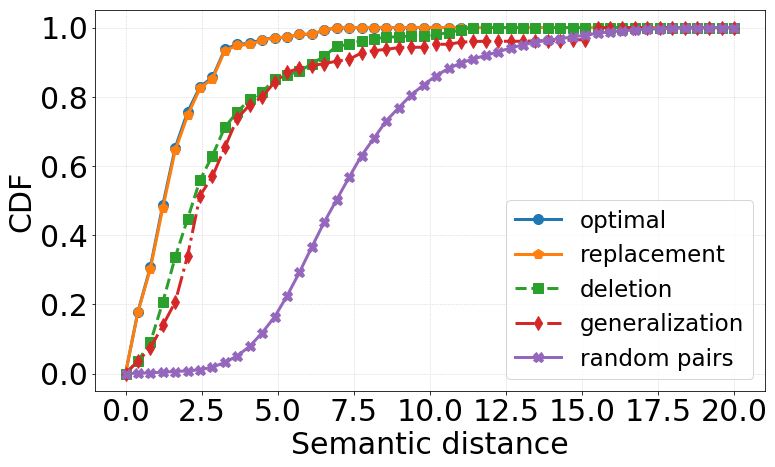}\label{fig:cdf_utility}}
\hspace{5mm}
\subfigure[]{\includegraphics[width=0.57\columnwidth]{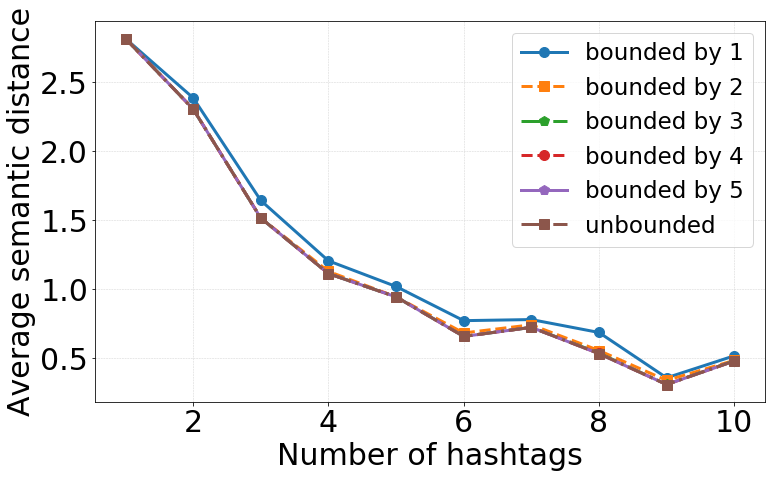}\label{utility}}
\hspace{5mm}
\subfigure[]{\includegraphics[width=0.57\columnwidth]{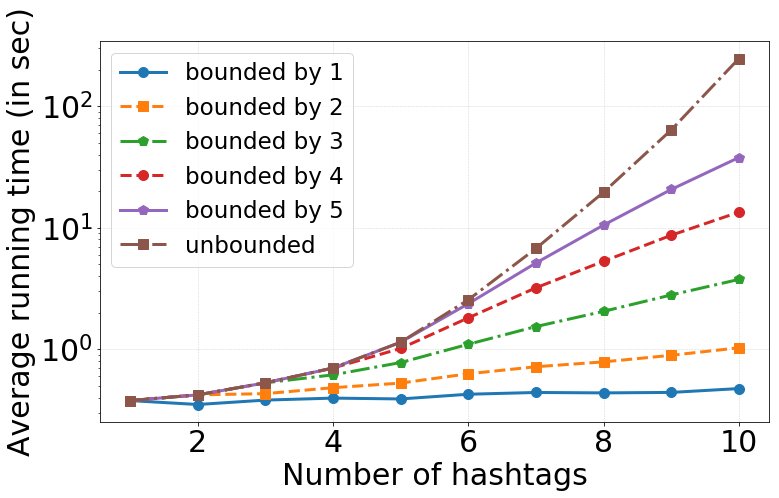}\label{time}}
\caption{(a) Cumulative distribution function (CDF) of the minimum utility loss, 
i.e., semantic distance, for maximum privacy constraint, 
of the three obfuscation mechanisms 
and the optimal one among all mechanisms. 
(b) Average minimum semantic distance of the original hashtags 
from the optimal solution (unbounded) 
and solutions with an upperbound on the number of hashtags 
to be obfuscated with respect to the number of original hashtags (x-axis) 
(c) Average running time of Tagvisor (per sample) with respect to the number of original hashtags.} 
\end{figure*}

\subsection{Utility vs. Privacy} 
While the previous subsection does not consider any optimization of utility, 
in this subsection we assume that the user relies on the optimization problem 
defined in~(\ref{equa:utility}). 
For our experiments, we measure location privacy by the inaccuracy and set $\alpha = 1$. 
This means that the user wants to release a subset $\Hashtag_\photo'$  
that leads to a misclassification of the location and minimizes the utility loss. 
Note that, for a very few samples in the testing set, 
the adversary can still infer the correct location 
based solely on the locations distribution (baseline in Section~\ref{sec:eval}), 
against which Tagvisor cannot help.

Figure~\ref{fig:cdf_utility} shows the cumulative distribution function 
of the minimal utility loss over all test samples, 
for the three obfuscation mechanisms, 
and the optimal one (when the optimization considers all mechanisms together). 
First, we observe that replacement provides close to optimal utility 
and is selected for the optimal solution in 85\% of the cases, 
against 14\% for hiding, and 1\% for generalization. 
We also notice that hiding provides higher utility than generalization 
if the mechanisms are considered separately. 
We explain this result as follows: 
First, replacement misleads the classifier 
but at the same time keeps enough utility 
as the fake hashtags are by design chosen to be semantically close 
to the original hashtags. 
Second, generalization provides worse utility 
than hiding because it must modify more hashtags 
than hiding to reach the same privacy level, thereby increasing the semantic distance. 

We further compare the absolute semantic distances 
given by all four approaches 
and the semantic distance in random pairs of posts (purple plain curve). 
This baseline utility curve is constructed 
by randomly sampling 10,000 pairs of posts in our dataset. 
We observe in Figure~\ref{fig:cdf_utility} that 
90\% of the privacy-preserving hashtag sets 
have a semantic distance smaller than 3 to the original hashtag sets for replacement whereas, 
for deletion and generalization, the corresponding distance is around 6, 
and the distance between random pairs goes up to 11. 
This demonstrates that we can keep a relatively fair amount of semantic utility with the replacement mechanism.

Finally, we observe in Figure~\ref{utility} (``unbounded'' curve) that 
the utility loss is in general smaller when the original number of hashtags is larger. 
This can be explained by the fact that, when we increase the number of hashtags, 
we have a larger set of hashtags to obfuscate. 
Obfuscating 2 or 3 hashtags already brings enough privacy, 
as shown in Figure~\ref{fig:accuracy_obfuscation}. 
In other words, the relative change in the semantics of hashtags 
is smaller when we have a larger original set of hashtags. 
When the user has originally one or two hashtags 
with which the adversary correctly infers his location, 
he has to remove one or two hashtags to mislead the adversary, and thus loses most of his utility.

\subsection{Time Performance}

We now present the running time which is an important factor for Tagvisor's usability, 
and we propose solutions for trading off time efficiency with optimality. 
All the experiments are conducted on a laptop computer with 3.3 GHz CPU and 16 GB memory, 
and are implemented relying on Python packages 
such as pandas, numpy, and scikit-learn.
It is worth noting that our experiments
can be further optimized by using more efficient programming languages or libraries, 
and by parallelizing the computations, 
e.g., of the three obfuscation methods. 
Also, current smartphones' computing capabilities 
are shown to be close to those of laptops \cite{iphone}.

Figure~\ref{time} shows how the running time evolves 
with an increasing number of hashtags. 
It clearly demonstrates that the optimal solution (``unbounded'' curve) 
is not practical when the original number of hashtags goes beyond 5. 
For instance, it takes more than 2 seconds for 6 hashtags. 
Even if most posts are shared with fewer than 6 hashtags, 
we want to provide a (nearly) optimal \emph{and} 
efficient mechanism for users willing to share 6 or more hashtags. 

Each curve in Figure~\ref{time} denotes a different maximum number of hashtags 
that can be obfuscated. 
We observe that, by bounding this maximum number to two, 
we obtain a total running time of around one second for any number of original hashtags, 
which is very satisfactory.
We can even reach half a second by bounding it to one hashtag. 
At the same time, we can observe in Figure~\ref{utility} 
that the utility obtained with this maximum number bounded by one 
is very close to the optimal utility already. 
With this number bounded by two, 
the corresponding utility curve and the optimal one can hardly be differentiated. 
From a privacy perspective, 
even by bounding to two obfuscated hashtags, 
we get an average accuracy equal to 0.038, 
which is smaller than the baseline value (0.053) reported in Section~\ref{sec:eval}. 
Only with an upperbound of one hashtag, 
we get an accuracy (equal to 0.21) greater than the baseline. 
This shows that bounding the number of possible hashtags 
to be obfuscated to two provides the best utility-privacy-efficiency trade-off.

\section{Discussion}
\label{sec:discussion}

We discuss here some important details of our proposal. 
First of all, our attack is orthogonal to the use of computer vision techniques 
as it does not rely on the photo content or any visual background at all. 
It relies only on hashtags to make it applicable to a wider variety of settings. 
Of course, if the posts contain photos in addition to hashtags and the background of the photos is comprehensible, 
a human or advances in computer vision might sometimes 
have better chances of identifying the location of a photo. 
The evaluation of the privacy impact of such cases is left for future work.

Second, one may think that the set of sanitized hashtags shared via Tagvisor 
could be used by the adversary to improve his machine-learning model. 
Despite being intuitive, this statement does not hold. 
First, the location is by definition not shared with the hashtags released using Tagvisor, 
thus cannot be used to train the adversary's classifier. 
Indeed, someone uses Tagvisor if he wants to hide his location. 
Moreover, the location that the adversary infers via hashtags does not provide him 
with any extra information than what is already contained in his trained classifier. 
Only the new knowledge brought by users sharing 
both hashtags and their locations (i.e., not using Tagvisor) 
will be incorporated into the knowledge base $K$. 

Third, our semantics-based utility metric does not 
necessarily encompass the purposes of all users. 
This motivated us to leave some degrees of freedom to the users in Tagvisor, 
as explained in Section~\ref{sec:defense}. 
Concretely, the user can decide to either let the application 
directly suggest the optimal set of hashtags, 
or suggest the best sets for each obfuscation mechanism. 
The user can even set a minimum/maximum number of hashtags to be shared, 
also for each mechanism, and receive recommendations. 
We believe that this approach brings both usability 
and maximal utility for all users. 
A user study to analyze the acceptance of Tagvisor
among OSN users would be an interesting future work.

Finally, by requiring that posts contain both hashtags and locations, 
our dataset may be biased towards users that are more willing to share locations.
Our high accuracy in the location inference attack 
could be partially attributed to this bias in the testing set. 
However, one must realize that we cannot avoid such a restriction 
as we need the ground truth on the location 
to evaluate the attack performance. 
In the worst case, our attack evaluation provides a lower bound 
on the average privacy of sharing hashtags. 
Moreover, as shown in Figure~\ref{fig:hash_dist}, 
users usually share hashtags, 
to a large extent regardless of whether they jointly share their location.
The training set is a priori not biased, 
as even the adversary must have access 
to both location and hashtag information to train his classifier.

\section{Related Work}
\label{sec:relwork}

\noindent \textbf{Location privacy. }
Shokri et al.~\cite{STBH11} propose a comprehensive framework for 
quantifying location privacy.
This framework is able to capture various prior information available to an attacker and
attacks the attacker can perform. 
The authors of~\cite{STBH11} 
also propose several privacy metrics,
which inspire the metrics used in this paper. 
A{\u{g}}{\i}r et al.~\cite{AHHH16} further analyze 
the impact of revealing location semantics, such as restaurant or cinema,
by extending the underlying graphical model used for location inference in~\cite{STBH11}.
Experimental results on a Foursquare dataset demonstrate 
that location semantics can raise severe privacy concerns.
Bilogrevic et al.~\cite{BHMSH15} 
concentrate on the utility implications of one popular privacy-preserving mechanism, 
namely generalization (both geographically and semantically).
Through the data collected from 77 OSN users,
their model is able to accurately predict the motivation behind location sharing,
which enables design of privacy-preserving location-sharing applications.
Other interesting works in this field include~\cite{OTP17,PTD17,PTD18}.
Contrary to these previous works, 
ours focuses on the location-privacy risks of hashtag sharing, and propose mechanisms to mitigate these risks.

\smallskip
\noindent \textbf{Hashtag analysis. } 
Highfield and Leaver \cite{HL14} point out the research challenges for Instagram, 
especially in comparison to Twitter, 
by examining hashtags in the two OSNs.
Based on a dataset collected from Instagram users' profiles and hashtags, 
Han et al. \cite{HLJJL16} identify teens and adults using popular supervised learning techniques 
and additionally reveal significant behavioral differences between the two age groups. 
The authors of~\cite{SCFYCQA15} use \#selfie in Instagram 
to study the phenomenal ubiquitous convention of self-portrait,
while the authors of~\cite{MAH16} study the health implications 
and obesity patterns associated with the use of \#foodporn. 
More recently, Lamba et al.~\cite{LBVAASK17} study \#selfie in Twitter that can potentially cause life threat.
Although no previous work has addressed privacy issues arising out of hashtags to the best of our knowledge, 
the aforementioned works have been very influential 
in shaping our approach of using hashtags for training our attacker model.

\smallskip
\noindent \textbf{Location prediction in OSNs. } 
Multiple works aim at estimating a user's home location using the posts he has published in OSNs. 
Cheng et al.~\cite{CCL10} identify words in tweets with a strong local geo-scope 
and estimate several possible locations for each user with decreasing probabilities. 
Chandra et al. \cite{CKM11} predict city-level user location 
relying on a communication model in Twitter's conversation. 
Jurgens et al.~\cite{JFMXR15} recently 
provide a survey on the related works in this direction.
Different from these studies, we address the problem of identifying the precise location of users' posts.

Some other works concentrate on estimating the location of each post, 
which is closer to the problem we address. 
Li et al.~\cite{LSVEL11} use KL-Divergence to build a language model for each location.
Due to the insufficient number of tweets for each location, 
they incorporate texts in web pages returned by search engines 
to support their model. 
Agerwal et al.~\cite{AVSS12} also use geographic gazeteer data 
as vocabulary to identify location names from phrases in tweets. 
The approach of Dalvi et al.~\cite{DKP12} 
assumes that a geo-located object is mentioned in the tweet. 
Our attack however does not need any external knowledge 
and already shows a high accuracy in predicting the exact point of interest. 
Moreover, we propose a tool for privacy-preserving hashtag sharing.

\section{Conclusion}
\label{sec:conclu}

The rapid adoption of new concepts and tools poses novel risks 
towards privacy that users are not always aware of. 
In this paper, we thoroughly investigate the impact of hashtag sharing on location privacy. 
We present and show the effectiveness of a random forest classifier 
that uses hashtags for fine-grained location inference within and across cities. 
In order to counter the proposed attack, 
we propose three obfuscation mechanisms 
and an efficient system that determines the optimal/near-optimal set of hashtags 
to preserve privacy without degrading significantly the original utility. 

Our work demonstrates a clear threat towards location privacy stemming from the use of hashtags. 
Even though some users intentionally use hashtags 
to reveal their (more or less accurate) locations, 
many of them do not make use of hashtags 
that are clearly linked to location information, 
and thus should expect some level of location privacy. 
In the context of privacy, 
even a moderately large absolute number of affected individuals 
represents a serious enough negative effect~\cite{CBCSHK10}, 
which is effectively reduced by our proposed privacy-preserving system, Tagvisor.

As future work, we plan to study how the publication time 
of the post can help improve the classifier's success. 
Besides temporal information, 
other cues could boost the accuracy of the inference attack, 
for example, social proximity to already geo-located users 
or use of deep learning on the images to identify background locations, etc. 
All these can further degrade users' location privacy, 
and it will be interesting to incorporate them in our framework in the future. 
Another direction for future work would be 
to incorporate defenses against other privacy threats arising out of hashtags 
such as demographics, linkability and friendship between users.

\begin{acks}
This work was partially supported by the German Federal Ministry of Education and
Research (BMBF) through funding for the Center for IT-Security,
Privacy and Accountability (CISPA) (FKZ: 16KIS0656).
The authors would like to thank Rose Hoberman
for her valuable comments on the submitted manuscript. 
\end{acks}

\bibliographystyle{ACM-Reference-Format}
\bibliography{tag2loc}
\end{document}